\documentclass[a4paper,twoside,10pt]{article}

\input{jgrg19.sty}
\begin{document}
%
\pagestyle{fancy}
\fancyhead{}
  \fancyhead[RO,LE]{\thepage}
  \fancyhead[LO]{T.Igata}                 
  \fancyhead[RE]{Toroidal Spiral Strings in Higher-dimensional Spacetime}    
\rfoot{}
\cfoot{}
\lfoot{}
\label{P12}    
\title{%
  Toroidal Spiral Strings in Higher-dimensional Spacetime
}
\author{%
  Takahisa Igata\footnote{E-mail:igata@sci.osaka-cu.ac.jp}
  and 
  Hideki Ishihara\footnote{E-mail:ishihara@sci.osaka-cu.ac.jp}
}
\address{%
  Department of Mathematics and Physics, 
  Graduate School of Science, \\
Osaka City University,  Osaka 558--8585, Japan
}
\abstract{
We report on our progress in research of separability of the Nambu-Goto equation for test strings with a symmetric configuration in a shape of toroidal spiral in a five-dimensional Kerr-AdS black hole. In particular, for a \lq{\it Hopf loop}\rq\ string which is a special class of the toroidal spirals, 
we show the complete separation of variables occurs in two cases, Kerr background and Kerr-AdS background with equal angular momenta. 
We also obtain the dynamical solution for the Hopf loop around a black hole 
and for the general toroidal spiral in Minkowski background. 
}

\section{INTRODUCTION}
Recently, much attention has been focused on the study of higher-dimensional spacetime. 
One of our important task is revealing properties of higher-dimensional black hole 
because identification of the spacetime dimension could be done by observations of phenomena concerning to black holes.

When we study black hole physics, a test particle plays an crucial role as a probe of black hole spacetime because it gives us information of the geometry around it. 
In addition a test string would be also a powerful tool to understand the black hole physics.

Here we discuss dynamics of a test string around a higher-dimensional spacetime. As is known that Kerr-AdS black hole spacetimes are typical exact solutions of the Einstein equation in arbitrary dimension. 
Now we consider the five-dimensional one, which corresponding metric is given by
\begin{eqnarray}
	ds^2
 		= - \frac{\Delta_{\theta} \Xi_r dt^2 }{\Xi_a \Xi_b}
 		+ \frac{2 M}{\Sigma}  \Big(\frac{\Delta_{\theta} dt}{\Xi_a \Xi_b}
 		- \nu \Big)^2
		+ \frac{\Sigma dr^2}{\Delta_r}
 + \frac{\Sigma d\theta^2}{\Delta_{\theta}}
 + \frac{r^2 + a^2}{\Xi_a} \sin^2\theta d\Phi^2
 + \frac{r^2 + b^2}{\Xi_b} \cos^2\theta d\Psi^2,
\end{eqnarray}
with
\begin{eqnarray}
	\Delta_r &=& \frac{(r^2 + a^2)(r^2 + b^2)(1 + \lambda^2 r^2)}{r^2} - 2 M ,\qquad
	\Delta_{\theta} = 1 - a^2 \lambda^2 \cos^2\theta
		 - b^2 \lambda^2 \sin^2\theta\ ,
\\
	\nu &=& a \sin^2\theta \frac{d\Phi}{\Xi_a} 
		+ b \cos^2\theta  \frac{d\Psi}{\Xi_b} ,
\qquad
	\Sigma = r^2 + a^2 \cos^2\theta + b^2 \sin^2\theta,
\\
	\Xi_a &=& 1 - a^2 \lambda^2, 
\qquad
	\Xi_b = 1 - b^2 \lambda^2 ,
\qquad
	\Xi_r = 1 + \lambda^2 r^2,
\end{eqnarray}
where $M$ is the mass parameter, and $a$ and $b$ are two independent rotational  parameters. The parameter $\lambda$ is connected with the cosmological 
constant $\Lambda$ as $\lambda^2=- \Lambda/6$. 
We should note that the spacetimes have a remarkable common property in arbitrary dimension, that is, separability in the geodesic Hamilton-Jacobi equation \cite{P12_Carter:1968ks, P12_Frolov:2003en, P12_Frolov:2006pe}. 
It is expected that a string motion could be also separable due to the geometrical property. 
In fact, separability is also shown for a string in a class of stationary string in \cite{P12_Frolov:2004qw, P12_Kubiznak:2007ca, P12_Ahmedov:2008pq}. 
Hence, we concentrate on the dynamical string in the five-dimensional Kerr-AdS black hole.

We discuss a special string which has a symmetry of configuration. 
We assume that one of the Killing vector fields, say $\xi$, on a target spacetime is tangent to a worldsheet of the string. 
We call it a {\it cohomogeneity-one string} associated with $\xi$. 
A stationary string is in a class of a cohomogeneity-one string where the Killing vector is timelike. 
It is known that the Nambu-Goto action for a cohomogeneity-one string reduces to a geodesic action in a quotient space ${\cal O}$, constructed by the isometry group generated by $\xi$, with a norm weighed metric $F h_{\mu \nu}$ as follows, 
\begin{eqnarray}
S_{NG}
\,=\,- \mu\,\Delta \sigma \int_{{\cal C}}\,
\sqrt{-\,F\,h_{\mu\nu}\,\frac{dx^{\mu}}{d\tau}\,\frac{dx^{\nu}}{d\tau}}d\tau, 
\end{eqnarray}
where $\mu$ is a tension of string, $\tau$ and $\sigma$ are the coordinates of the string worldsheet, ${\cal C}$ is a curve on ${\cal O}$, $F$ is a norm of $\xi$, and $h_{\mu\nu}$ is natural projection metric of ${\cal O}$ defined as $h_{\mu\nu} = g_{\mu \nu} - \xi_{\mu} \xi_{\nu}/F$.
Therefore, the problem for finding solution of motion of cohomogeneity-one strings associated with $\xi$ in the five-dimensional Kerr-AdS black hole reduces to the problem for solving geodesic equations in a four-dimensional space $({\cal O},  F h_{\mu \nu})$.

\begin{figure}[t]
\centering
 \includegraphics[width=8cm]{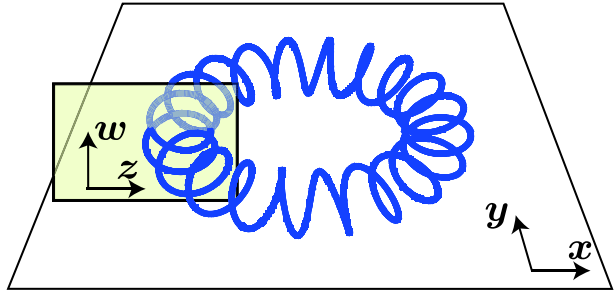}
\caption{A toroidal spiral string coils around a torus embedded in the four-dimensional space on a snap shot. 
}
\label{fig:P12_ToroidalSpiral}
\end{figure}

We concentrate on dynamics of a class of cohomogeniety-one strings, {\it toroidal spiral strings} \cite{P12_Igata:2009dr,P12_BlancoPillado:2007iz}. 
The string is associated with the Killing vector field 
\begin{eqnarray}
\xi = \partial_{\Phi} + \alpha \partial_{\Psi},
\end{eqnarray}
where $\Psi$ and $\Phi$ are correspond to the azimuthal angles on the independent two-dimensional planes in the four-dimensional spatial section and $\partial _{\Phi}$ and $\partial _{\Psi}$ are the commutable rotational Killing vector fields. 
The constant $\alpha$ describes winding ratio of the string. 
The string have a spiral shape along a circle on a time slice as shown in Figure \ref{fig:P12_ToroidalSpiral}.

We stress that the special class of the toroidal spiral strings with $\alpha^2=1 $ behaves "good" as discussed later. 
The string in this class lies along a fiber of Hopf fibration $S^3$ which is a constant surface of the radial coordinate on a timeslice in the Kerr-AdS black hole, after which we name the string as a {\it Hopf loop}. 
We will see that the Hopf loop has a special nature in the following section.

\section{SEPARABILITY}
Let us discuss the separability in the geodesic equation in $({\cal O}, F h_{\mu\nu})$ for a toroidal spiral string. In order to solve the geodesic equation, we use the Hamilton-Jacobi method. The Hamilton-Jacobi equation in our case is written by
\begin{eqnarray}
&&\bigg[\frac{(r^2 + a^2)(r^2 + b^2)\Xi_a \Xi_b - 2 M (r^2 + a^2 b^2 \lambda^2)}{r^2 \Delta_r}
 + \frac{2 (1 - a^2 \lambda^2) (1 - b^2 \lambda^2)}{\lambda^2 (a^2 + b^2) + \lambda^2 (a^2 - b^2) \cos2\theta - 2}\bigg] \frac{E^2}{\lambda^2}\nonumber
\\[0.2cm]
&& + 
\bigg[
4 a b M \alpha \Xi_r
 + (\alpha^2 - 1)(b^2 - a^2) r^2 \Xi_r
 + a^2 b^2 \Xi_r 
\Big[
(\alpha^2 - 1)(1 + \lambda^2 (a^2 - b^2)) - 2 \alpha^2
\Big]\nonumber
\\ [0.2cm]
&& + a^2 
(1 + \alpha^2 r^2 \lambda^2)(a^2 \Xi_r - 2 M)
 + b^2 
(\alpha^2 + r^2 \lambda^2)(b^2 \Xi_r - 2 M)
 + \Big(
\frac{\alpha^2 \Xi_a}{\sin^2\theta} + \frac{\Xi_b}{\cos^2 \theta}
\Big)
\bigg] L^2\nonumber
\\ [0.2cm]
	&& + \frac{4 M[- b (r^2 + a^2)
 		+ \alpha a (r^2 + b^2)]}{r^2 \Delta_r} E L
 	+ \Delta_r \Big(\frac{d S_r}{d r}\Big)^2
 	+ \Delta_{\theta}\Big(\frac{d S_{\theta}}{d \theta}\Big)^2
 	= -\mu^2 F \Sigma, 
\label{eqn:P12_Hamilton-Jacobi}
\end{eqnarray}
where $S_r$ and $S_{\theta}$, each term of the Hamilton's principal function, are functions $r $ and $\theta$, respectively, and $E$ and $L$ are conserved quantities correspond to the energy and the angular momentum of the string. 
We find that separation of variables does not occur for a general toroidal spiral string because the right-hand side of \eqref{eqn:P12_Hamilton-Jacobi}, which has the explicit form
\begin{eqnarray}
	\mu^2 F \Sigma
	&=& \mu^2 (r^2 + a^2 \cos^2\theta + b^2 \sin^2\theta)
	 \Big[\frac{(r^2 + b^2)\alpha^2 \cos^2\theta}{\Xi_b} 
		+ \frac{(r^2 + a^2) \sin^2\theta}{\Xi_a} \Big]
\nonumber\\ 
	&& + 2 M \mu^2 
	\Big(
	\frac{\alpha b \cos^2\theta}{\Xi_b} 
	+ \frac{a \sin^2\theta}{\Xi_a}\Big)^2,
\label{eqn:P12_mass_term}
\end{eqnarray}
does not allow the separation of variables. The complete separability in the Hamilton-Jacobi equation depends on the parameter $\alpha$, and parameters of the background geometry. 

When we consider a Hopf loop, $\alpha=1$, around the black hole, we see that the Hamilton-Jacobi equation can be separable for two cases: 
\begin{description}
\item[(A)] vanishing cosmological constant {\it i.e.}, the background is a Kerr black hole, 
\item[(B)] the black hole with non-zero cosmological constant and two equal angular momenta.
\end{description}
The complete separability implies that the metric $F h_{\mu\nu}$ admits Killing tensor fields. 
One obtains the irreducible and reducible Killing tensor field on the quotient space $({\cal O}, Fh_{\mu \nu})$ in the case of (A) and (B), respectively(see ref.\cite{P12_Igata:2009fd} for detail discussion).

\section{DYNAMICS}
Let us see dynamics of a Hopf loop around the Kerr-AdS black hole in the case of (B). 
The metric $Fh_{\mu\nu}$ has SO(3) symmetry because the base space of the Hopf bundle becomes round $S^2$, and $F$ is a function of $r$. 
Therefore, without loss of generality, we can restrict our attention to study geodesics confined in the equatorial plane, {\it i.e.}, $\theta=\pi/4$.

\begin{figure}[htbp]
 \begin{center}
 \includegraphics[width=70mm]{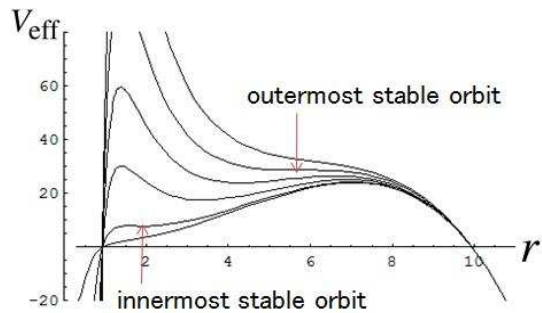}
 \end{center}
 \caption{
The effective potentials for radial motion of a Hopf loop with respect to each $L$ in the five-dimensional Kerr-AdS black hole with equal angular momenta. 
The parameter choice is $r_g = 2 M = 1,\  \Lambda/6=- \lambda^2 = 0.01,$ and $a = 1/4$.}
\label{fig:P12_effective_potential}
\end{figure}

Then, with appropriate choice of the Lagrange multiplier $N$, the radial equation of motion becomes $\dot r^2+V_{{\rm eff}}=E^2$, with the effective potential given by 
\begin{eqnarray}
 V_{{\rm eff}} = \frac{\mu^2 r^2 \Delta_r}{(r^2 + a^2)\Xi_a^2}
  + \frac{4 r^2 \Delta_r \Xi_a L^2}{[(r^2+a^2)^2\Xi_a + 2 M a^2] (r^2 + a^2)}.
\label{eq:potential}
\end{eqnarray}
Typical shapes of the effective potential for the case (B) are given in Figure \ref{fig:P12_effective_potential}. 
The figure shows that the radial motion of the Hopf loop is classified into two types, bounded or unbounded. 
This nature is understood that the motions of Hopf loops are driven by the three forces; tension of string, centrifugal force, and gravitational force, 
and the orbits are determined by the competition of these forces \cite{P12_Igata:2009dr}. 
The existence of bounded orbits for the Hopf loop is analogous to the case of a free particle around a four-dimensional black hole. 
We note that this is particular nature of the string because there is no bounded orbit for test particles around the five-dimensional Kerr black hole \cite{P12_Frolov:2003en}.

Stationary Hopf loop solution exists at the local minimum of $V_{\rm eff}$. 
By the effect of the gravitational force, there exists a critical radius of  Hopf loop for each black hole such that no stable Hopf loop is inside the radius, namely, the innermost stable orbit exists. 
In addition, in the case of $\Lambda > 0$, Hopf loops can grow up to infinite radius by the de Sitter expansion, and the outermost stable orbit exists. 

We have also studied dynamics of a toroidal spiral string in the five-dimensional Minkowski background and shown that
\begin{description}
\item[]- For a general toroidal spiral, {\it i.e.}, for all $\alpha$, the Hamilton-Jacobi equation is completely separable
\end{description}
 In addition, we can obtain the general solution which describes harmonic oscillations of radii of torus which is coiled by the toroidal spiral explicitly (see ref.\cite{P12_Igata:2009fd} for detail discussion).

\section{SUMMARY}
 In this proceeding, we have discussed the separability of the Nambu-Goto equations for a toroidal spiral string in the five-dimensional spacetime. 
We have found that the equation admits the separation of variables for a Hopf loop in a Kerr black hole and a Kerr-AdS black hole with two equal angular momenta.

We have also shown the dynamical properties of the Hopf loop strings in the five-dimensional black hole. There exist bounded orbits and unbounded orbits in the quotient space, and also exist the innermost or outermost stable orbit. 
Since the motions of Hopf loops are driven by three forces: tension of string, 
the centrifugal force, and force of gravity, the orbits are determined by the competition of these forces. 
The stationary configurations are achieved by the balance of these forces.

The existence of the bounded orbits of the Hopf loops around 
five-dimensional black holes makes us recall the free particles 
around a four-dimensional black hole. 
These results shows that the toroidal spiral string is a candidate of observational probe in place of a test particle in higher-dimensional spacetime. 
The existence of the stationary configuration and the bounded orbits of the string shows that there are prospects for long life of the closed string in the higher-dimensional universe.

\end{document}